\documentclass[prd,twocolumn,aps,superscriptaddress,nofootinbib,showpacs]{revtex4-1}

\usepackage[utf8]{inputenc}

\usepackage[justification=raggedright]{caption}

\usepackage{subfig}
\usepackage{dcolumn}
\usepackage{bm}
\usepackage{color}
\usepackage{soul,xcolor}
\usepackage{times}
\usepackage{hyperref}
\hypersetup{
  colorlinks=true,
  citecolor=blue,
  linkcolor=blue,
  urlcolor=blue}
\usepackage{epsfig}
\usepackage{dcolumn}

\usepackage{amsfonts}
\usepackage{amsmath}
\usepackage{amssymb}

\def\be{\begin{equation}}
\def\ee{\end{equation}}
\def\bea{\begin{eqnarray}}
\def\eea{\end{eqnarray}}
\def\gsim{\ \rlap{\raise 2pt\hbox{$>$}}{\lower 2pt \hbox{$\sim$}}\ }
\def\lsim{\ \rlap{\raise 2pt\hbox{$<$}}{\lower 2pt \hbox{$\sim$}}\ }
\def\dslash{\kern-4pt \not{\hbox{\kern-2pt $\partial$}}}
\def\pslash{\not{\hbox{\kern-2pt p}}}

\newcommand{\dcp}{\delta_{CP}}
\newcommand{\nova}{NO$\nu$A{}}

\setstcolor{blue}

\begin{document}
\DeclareGraphicsExtensions{.eps,.ps}


\title{Study of parameter degeneracy and hierarchy sensitivity of NO$\nu$A in presence of sterile neutrino}

\author{Monojit Ghosh}
\email[Email Address: ]{monojit@phys.se.tmu.ac.jp}
\homepage[\\ORCID ID: ]{http://orcid.org/0000-0003-3540-6548}
\affiliation{Department of Physics, Tokyo Metropolitan University, Hachioji, Tokyo 192-0397, Japan}

\author{Shivani Gupta}
\email[Email Address: ]{shivani.gupta@adelaide.edu.au}
\homepage[\\ORCID ID: ]{http://orcid.org/0000-0003-0540-3418}
\affiliation{Center of Excellence for Particle Physics at the Terascale (CoEPP), University of Adelaide, Adelaide SA 5005, Australia}

\author{Zachary M. Matthews}
\email[Email Address: ]{zachary.matthews@adelaide.edu.au}
\homepage[\\ORCID ID: ]{http://orcid.org/0000-0001-8033-7225}
\affiliation{Center of Excellence for Particle Physics at the Terascale (CoEPP), University of Adelaide, Adelaide SA 5005, Australia}

\author{Pankaj Sharma}
\email[Email Address: ]{pankaj.sharma@adelaide.edu.au}
\homepage[\\ORCID ID: ]{http://orcid.org/0000-0003-1873-1349}
\affiliation{Center of Excellence for Particle Physics at the Terascale (CoEPP), University of Adelaide, Adelaide SA 5005, Australia}

\author{Anthony G. Williams}
\email[Email Address: ]{anthony.williams@adelaide.edu.au}
\homepage[\\ORCID ID: ]{http://orcid.org/0000-0002-1472-1592}
\affiliation{Center of Excellence for Particle Physics at the Terascale (CoEPP), University of Adelaide, Adelaide SA 5005, Australia}

\begin{abstract}
The first hint of the neutrino mass hierarchy is believed to come from the long-baseline experiment NO$\nu$A. Recent results from the NO$\nu$A shows a mild preference towards the CP phase $\delta_{13} = -90^\circ$ and normal hierarchy. Fortunately this is the favorable area of the parameter space which does not suffer from the hierarchy-$\delta_{13}$ degeneracy and thus NO$\nu$A can have good hierarchy sensitivity for this true combination of hierarchy and $\delta_{13}$. Apart from the hierarchy-$\delta_{13}$ degeneracy there is also the octant-$\delta_{13}$ degeneracy. But this does not affect the favorable parameter space of NO$\nu$A as this degeneracy can be resolved with a balanced neutrino and antineutrino run. However, ff we consider the existence of a light sterile neutrino then there may be additional degeneracies which can spoil the hierarchy sensitivity of NO$\nu$A even in the favorable parameter space. In the present work we find that apart from the degeneracies mentioned above, there are additional hierarchy and octant degeneracies that appear with the new phase $\delta_{14}$ in the presence of a light sterile neutrino in the eV scale. In contrast to the hierarchy and octant degeneracies appearing with $\delta_{13}$, the parameter space for hierarchy-$\delta_{14}$ degeneracy is different in neutrinos and antineutrinos though the octant-$\delta_{14}$ degeneracy behaves similarly in neutrinos and antineutrinos. We study the effect of these degeneracies on the hierarchy sensitivity of NO$\nu$A for the true normal hierarchy.
\end{abstract}
\maketitle

\section{Introduction}
Neutrino oscillation physics has developed significantly since its discovery, with precision measurements finally being carried out for the mixing parameters. In the standard three flavor scenario, neutrino oscillation is parametrized by three mixing angles: $\theta_{12}$, $\theta_{23}$ and $\theta_{13}$, two mass squared differences: $\Delta m^2_{21}$ and $\Delta m^2_{31}$ and one Dirac type CP phase $\delta_{13}$. Among these parameters the current unknowns are: (i) the sign of $\Delta m^2_{31}$ which gives rise to two possible orderings of the neutrinos which are: normal ($\Delta m^2_{31} >$ 0 or NH) and inverted ($\Delta m^2_{31} < 0$ or IH) (ii), two possible octants of the mixing angle $\theta_{23}$ which are lower ($\theta_{23} < 45^\circ$ or LO) and higher ($\theta_{23} > 45 ^\circ$ or HO), and (iii) finally the phase $\delta_{13}$. The currently running experiments intending to discover these unknowns are T2K \cite{Abe:2017uxa} in Japan and NO$\nu$A \cite{Adamson:2017gxd} at Fermilab. The main problem in determining the oscillation parameters in long-baseline experiments is the existence of parameter degeneracy \cite{Barger:2001yr,Ghosh:2015ena}. Parameter degeneracy implies same value of oscillation probability for two different sets of oscillation parameters. In standard three flavor scenario, currently there are two types of degeneracies: (i) hierarchy-$\delta_{13}$ degeneracy \cite{Prakash:2012az} and (ii) octant-$\delta_{13}$ degeneracy \cite{Agarwalla:2013ju}. The dependence of hierarchy-$\delta_{13}$ degeneracy is same in neutrinos and antineutrinos but the octant-$\delta_{13}$ degeneracy behaves differently for neutrinos and antineutrinos \cite{Ghosh:2014zea,Ghosh:2015tan}. Thus the octant-$\delta_{13}$ degeneracy can be resolved with a balanced run of neutrinos and antineutrinos 
but a similar method cannot remove the hierarchy-$\delta_{13}$ degeneracy. However, despite the hierarchy-$\delta_{13}$ degeneracy being unremovable in general, the parameter space can be divided into a favorable region where it is completely absent for long-baseline experiments, and an unfavorable region where it is present. For NO$\nu$A, the favorable parameter space is around \{NH, $\delta_{13} = -90^\circ$\} and \{IH, $\delta_{13} = +90^\circ$\} whereas the unfavorable parameter space is around \{NH, $\delta_{13} = 90^\circ$\} and \{IH, $\delta_{13} = -90^\circ$\}. The recent data from NO$\nu$A shows a mild preference towards $\delta_{13} = -90^\circ$ and normal hierarchy \cite{Adamson:2017gxd}. From the above discussion we understand that for these combinations of true hierarchy and true $\delta_{13}$, NO$\nu$A can have good hierarchy sensitivity and thus it is believed that the first evidence for the neutrino mass hierarchy will come from the NO$\nu$A experiment. However the understanding of degeneracies can completely change in new physics scenarios. This occurs for example if there exists a light sterile neutrino in addition to the three active neutrinos (the 3+1 scenario).

Sterile neutrinos are SU(2) singlets that do not interact with the Standard Model (SM) particles but can take part in neutrino oscillations. 
Recently there has been some experimental evidence supporting the existence of a light sterile neutrino at the eV scale. This has motivated re-examination of oscillation analyses of the long-baseline experiments in the presence of sterile neutrinos 
\cite{Hollander:2014iha,Agarwalla:2016mrc,Agarwalla:2016xxa,Agarwalla:2016xlg,Palazzo:2015gja,Gandhi:2015xza,Dutta:2016glq,Klop:2014ima,Bhattacharya:2011ee,Berryman:2015nua,Kelly:2017kch}. For details regarding the first hints of the existence of sterile neutrinos and for the current status we refer to Refs 
\cite{Adamson:2011ku,Abazajian:2012ys,Palazzo:2013me,Lasserre:2014ita,An:2014bik,Ade:2015xua,Gariazzo:2015rra,An:2016luf,Choubey:2016fpi,Aartsen:2017bap,Kopp:2013vaa,Ko:2016owz}. In the presence of an extra sterile neutrino, there will be three new mixing angles namely $\theta_{14}$, $\theta_{24}$ and $\theta_{34}$, two new Dirac type CP phases $\delta_{14}$, $\delta_{34}$ and one new mass squared difference $\Delta m^2_{41}$. Thus in the presence of these new parameters there can be additional degeneracies involving the standard mixing parameters and sterile mixing parameters. In this work we study the parameter degeneracy in this increased parameter space in detail. From the probability level analysis we find that in 3+1 case, we have two new kind of degeneracies which are the (i) hierarchy-$\delta_{14}$ and (ii) octant-$\delta_{14}$ degeneracies. Our results also show that in this case the scenario is completely opposite to that of the hierarchy and octant degeneracy arising with $\delta_{13}$. The hierarchy-$\delta_{14}$ degeneracy is opposite for both neutrinos and antineutrinos but the octant-$\delta_{14}$ degeneracy behaves similarly in neutrinos and antineutrinos. Thus unlike the octant-$\delta_{13}$ degeneracy, the octant degeneracy in this case can not be resolved by a combination of neutrino and antineutrino runs while the hierarchy degeneracy can be resolved with a balanced combination of neutrino and antineutrinos which was not the case for the hierarchy-$\delta_{13}$ degeneracy. To show the degenerate parameter space in terms of $\chi^2$, we present our results in the  $\theta_{23}({\rm test})$-$\delta_{13}({\rm test})$ plane taking different values of $\delta_{14}$. We do this for two values of $\theta_{23}$ (true): one in LO and one in HO and for the current best-fit of NO$\nu$A i.e., $\delta_{13} = -90^\circ$ and NH (favorable parameter space). We show this for considering (i) NO$\nu$A running in pure neutrino mode and (ii) NO$\nu$A running in equal neutrino and equal antineutrino mode. Next we discuss the effect of these degeneracies on the hierarchy sensitivity of NO$\nu$A. We find that because of the existence of hierarchy-$\delta_{14}$ and octant-$\delta_{14}$ degeneracy, the hierarchy sensitivity of NO$\nu$A is highly compromised at the current best-fit value of NO$\nu$A (i.e. $\delta_{13} = -90^\circ$ and NH). To show this we plot hierarchy sensitivity of NO$\nu$A in the $\theta_{23}$ (true)-$\delta_{13}$ (true) plane taking different true values of $\delta_{14}$ for NH. We also identify the values of $\delta_{14}$ for which the hierarchy sensitivity of NO$\nu$A gets affected. To the best of our knowledge this is the first comprehensive analysis of parameter degeneracies and their effect on hierarchy sensitivity in presence of a sterile neutrino has been carried out.

The structure of the paper goes as follows. In Section \ref{sec2} we discuss the oscillatory behaviour in the 3+1 neutrino scheme. In Section \ref{sec3} we	 give our experimental specification. In section \ref{sec4} we discuss the the various degeneracies both at probability and event level. In Section \ref{sec5} we give our results for hierarchy sensitivity and finally in Section \ref{sec6} we present our conclusions.

\section{Oscillation Theory}
\label{sec2}

The PMNS matrix can be parametrized in many ways, the most common form with three neutrino flavors is:
\begin{equation}
U_{\mathrm{PMNS}}^{3\nu}
=
U(\theta_{23},0)
U(\theta_{13},\dcp)
U(\theta_{12},0)\,.
\end{equation}where $U(\theta_{ij},\delta_{ij})$ contains a corresponding $2\times2$ mixing matrix:
\begin{equation}
U^{2\times 2}(\theta_{ij},\delta_{ij})
=
\left(
\begin{array}{c c}
\mathrm{c}_{ij} & \mathrm{s}_{ij}e^{i\delta_{ij}}\\
-\mathrm{s}_{ij}e^{i\delta_{ij}} & \mathrm{c}_{ij}
\end{array}
\right)
\end{equation}embedded in an $n\times n$ array in the $i,j$ sub-block. Note the abbreviation of trigonometric terms:
\begin{align}
\mathrm{s}_{ij}=&\sin\theta_{ij},\\
\mathrm{c}_{ij}=&\cos\theta_{ij}.
\end{align}We also use the conventions for mass-squared differences
\begin{equation}
\Delta m^2_{ij}=m^2_i-m^2_j\,,
\end{equation}
and we write the oscillation factors
\begin{align}
\Delta_{ij}=&\frac{\Delta m^2_{ij}L}{4E}\,.
\end{align}Extending to four flavors we use the parametrization:
\begin{equation}
U_{\mathrm{PMNS}}^{4\nu}=
U(\theta_{34},\delta_{34})
U(\theta_{24},0)
U(\theta_{14},\delta_{14})
U_{\mathrm{PMNS}}^{3\nu}\,.
\end{equation}
Where the three new matrices introduce the new mixing angles: $\theta_{14},\theta_{24},\theta_{34}$ and phases: $\delta_{14},\delta_{34}$. The final new oscillation parameter is the fourth independent mass-squared difference which comes into the probability and is usually chosen to be $\Delta m^2_{41}$ to remain consistent with the $3\nu$ parameters. Assuming that $\Delta m^2_{41}\gg\Delta m^2_{31}$, and that we are operating near the oscillation maximum where $\sin^2\Delta_{31}\approx1$, then the sterile-induced oscillations from $\sin^2\Delta_{41}$ terms will be very rapid. Hence the four flavor vacuum $\nu_\mu$ to $\nu_e$ oscillation probability can be averaged over the sterile oscillation factor $\Delta_{41}$ i.e.
\begin{align}
&\langle\sin^2\Delta_{41}\rangle
=
\langle\cos^2\Delta_{41}\rangle
=
\frac{1}{2}
\\
&\langle\sin\Delta_{41}\rangle
=
\langle\cos\Delta_{41}\rangle
= 0
\end{align}
this reflects the inherent averaging that the long-baseline detectors see due to the very short wavelength of the sterile induced oscillations and their limited energy resolution.

Once the averaging has been done the probability expression can be written using the conventions and approach from \cite{Klop:2015} as:
\begin{align}
	P^{4\nu}_{\mu e}
	=\quad & P^\mathrm{ATM}
	+P^\mathrm{SOL}
	+P^\mathrm{STR}\\\notag
	+&P^\mathrm{INT}_\mathrm{I}
	+P^\mathrm{INT}_\mathrm{II}
	+P^\mathrm{INT}_\mathrm{III},
\end{align}where $P^\mathrm{ATM}, P^\mathrm{SOL}$ and $P^\mathrm{INT}_\mathrm{I}$ are modified from the three flavor probability terms by the factor $(1-\mathrm{s}^2_{14}-\mathrm{s}^2_{24})$, i.e.
\begin{align}
	P^\mathrm{ATM}&=(1-\mathrm{s}^2_{14}-\mathrm{s}^2_{24})P^\mathrm{ATM}_{3\nu},\\
	P^\mathrm{SOL}&=(1-\mathrm{s}^2_{14}-\mathrm{s}^2_{24})P^\mathrm{SOL}_{3\nu},\\
	P^\mathrm{INT}_\mathrm{I}&=(1-\mathrm{s}^2_{14}-\mathrm{s}^2_{24})P_{3\nu}^\mathrm{INT}.
\end{align}
With the $3\nu$ terms:
\begin{align}
	P_{3\nu}^\mathrm{ATM}&\approx4\mathrm{s}^2_{23}\mathrm{s}^2_{13}\sin^2\Delta_{31},\\
	P_{3\nu}^\mathrm{SOL}&\approx4\mathrm{c}^2_{12}\mathrm{c}^2_{23}\mathrm{s}^2_{12}\sin^2\Delta_{21},\\
	P_{3\nu}^\mathrm{INT}&\approx8\mathrm{s}_{13}\mathrm{s}_{12}\mathrm{c}_{12}\mathrm{s}_{23}\mathrm{c}_{23}\sin\Delta_{21}\sin\Delta_{31}\cos(\Delta_{31}+\delta_{13}).
\end{align}

The new $4\nu$ terms are

\begin{align}
	P^\mathrm{STR}
	&\approx2\mathrm{s}^2_{14}\mathrm{s}^2_{24},\\
	P^\mathrm{INT}_\mathrm{II}
	&\approx4\mathrm{s}_{14}\mathrm{s}_{24}\mathrm{s}_{13}\mathrm{s}_{23}\sin\Delta_{31}\sin(\Delta_{31}+\delta_{13}-\delta_{14}),\\
	P^\mathrm{INT}_\mathrm{III}
	&\approx-4\mathrm{s}_{14}\mathrm{s}_{24}\mathrm{c}_{23}\mathrm{s}_{12}\mathrm{c}_{12}\sin(\Delta_{21})\sin\delta_{14}.
\end{align}
However, in the case of \nova{} we can simplify this with approximations. Again, from \cite{Klop:2015},the constraints on the sterile mixing angles imply that the absolute values for $P^\mathrm{SOL},\ P^\mathrm{STR}$ and $P^\mathrm{INT}_\mathrm{III}$ are less than 0.003 so can be neglected. Additionally, for simplicity we neglect the terms multiplied by $\mathrm{s}_{14}$ and $\mathrm{s}_{24}$ in $P^\mathrm{ATM}$ and $P^\mathrm{INT}$, leaving:
\begin{align}
	P^{4\nu}_{\mu e}
	\approx P^\mathrm{ATM}_{3\nu}
	+P^\mathrm{INT}_\mathrm{3\nu}
	+P^\mathrm{INT}_\mathrm{II}.
\end{align}
which is:
\begin{eqnarray}\label{eq1}
	P^{4\nu}_{\mu e}
	=
	&\quad 4\mathrm{s}^2_{23}\mathrm{s}^2_{13}\sin^2\Delta_{31}\\
	&+8\mathrm{s}_{13}\mathrm{s}_{12}\mathrm{c}_{12}\mathrm{s}_{23}\mathrm{c}_{23}\sin\Delta_{21}\sin\Delta_{31}\cos(\Delta_{31}+\delta_{13})\notag\\
	&+4\mathrm{s}_{14}\mathrm{s}_{24}\mathrm{s}_{13}\mathrm{s}_{23}\sin\Delta_{31}\sin(\Delta_{31}+\delta_{13}-\delta_{14})\notag.
\end{eqnarray} 
From the $\Delta_{31}, \delta_{13}$ and $\delta_{41}$ dependent terms arise the hierarchy-CP degeneracies, due to the unconstrained sign of $\Delta_{31}$ and the (mostly) unconstrained CP phases $\delta_{13}$ and $\delta_{14}$, which can compensate for sign changes in $\Delta_{31}$.
The above formula is for neutrinos. The relevant formula for the antineutrinos can be obtained by replacing $\delta_{13}$ by $-\delta_{13}$ and $\delta_{14}$ by $-\delta_{14}$. Note that the above expression is for vacuum and free from the parameters $\theta_{34}$ and $\delta_{34}$.

\begin{figure*}\centering
\includegraphics[scale=0.75]{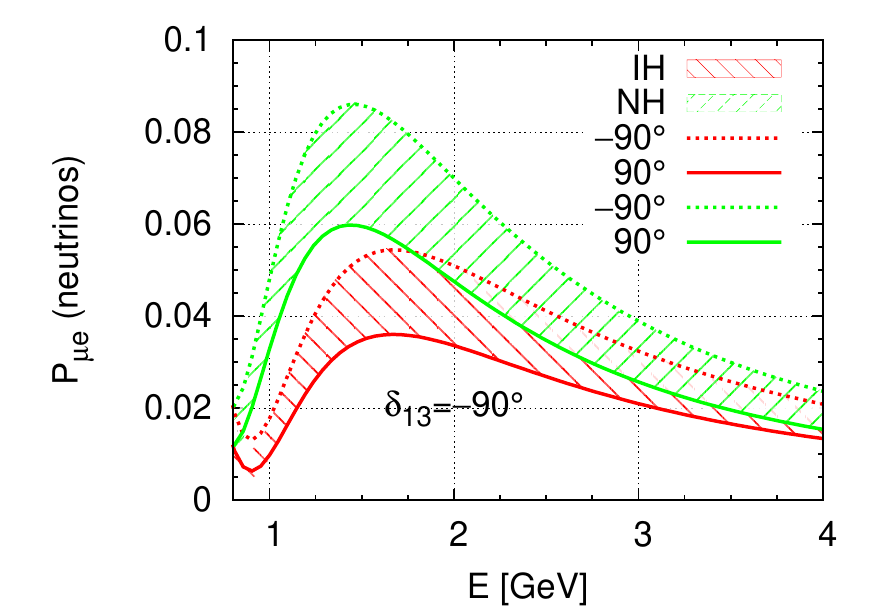}
\includegraphics[scale=0.75]{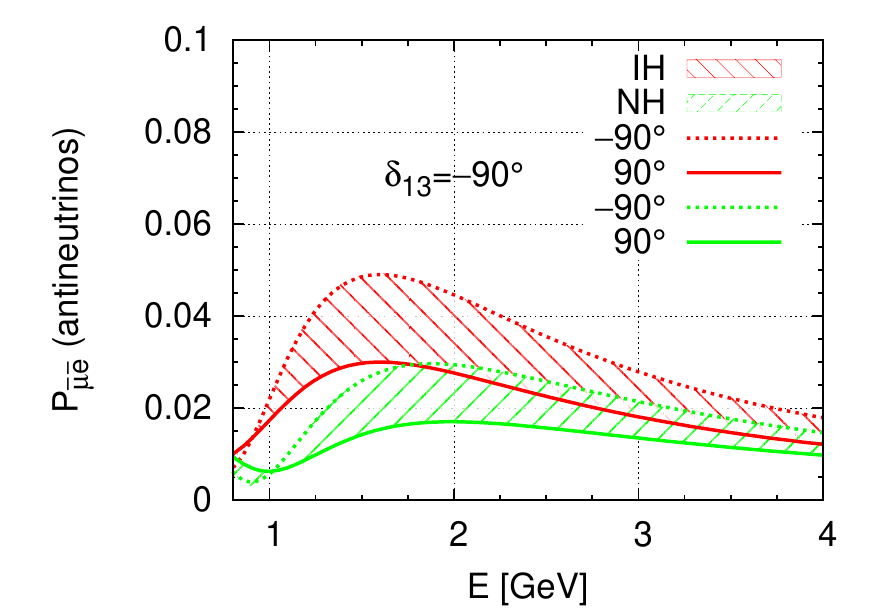}\\
\includegraphics[scale=0.75]{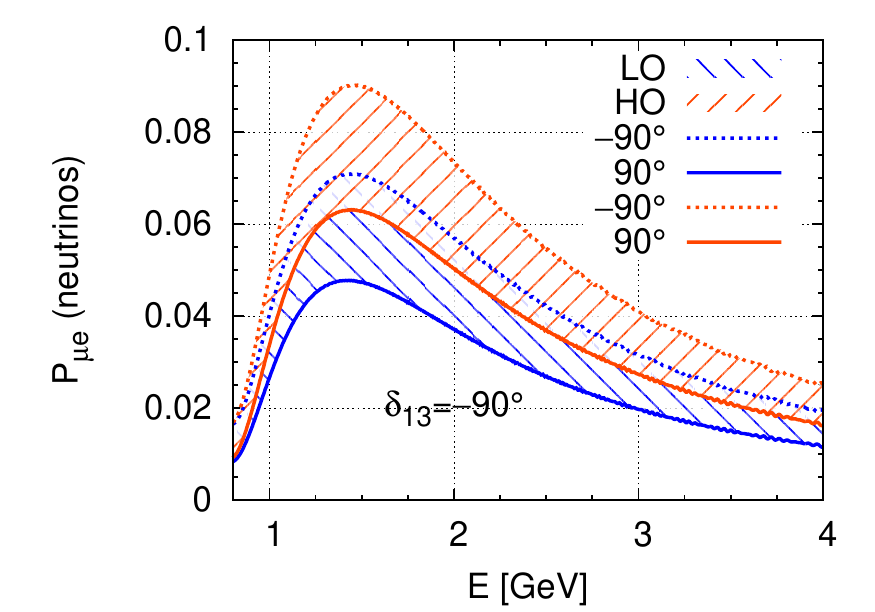}
\includegraphics[scale=0.75]{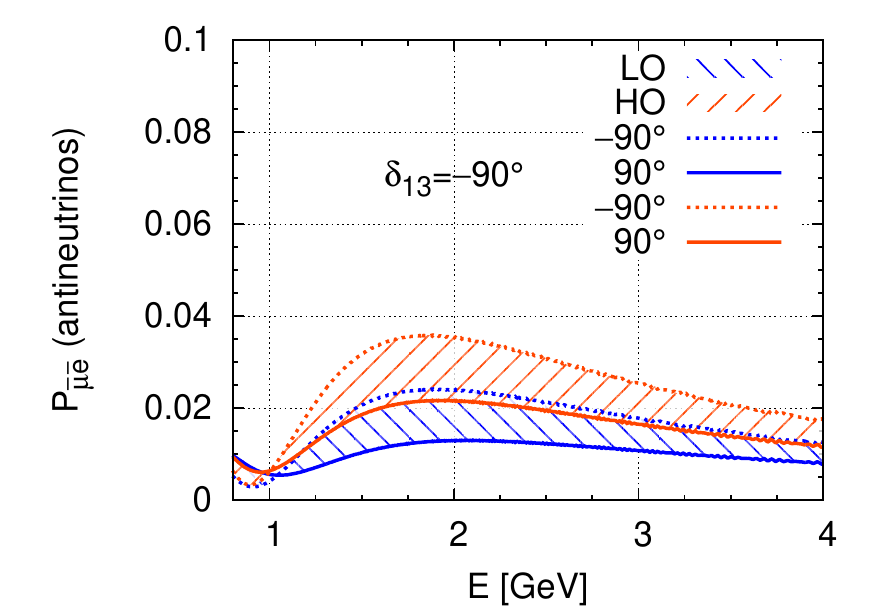}
\caption{$\nu_\mu \rightarrow\nu_e$ oscillation probability bands for $\delta_{13} = -90^\circ$. Left panels are for neutrinos and right panels are for antineutrinos. The upper panel shows the hierarchy-$\delta_{14}$ degeneracy and the lower panels shows the octant-$\delta_{14}$ degeneracy.}
\label{fig:prob}
\end{figure*}

\section{Experimental Specification}
\label{sec3}

For our analysis we consider the currently running long-baseline experiment \nova{}. \nova{} is an 812 km baseline experiment using the NuMI beam 
line at Fermilab directing a beam of $\nu_\mu$'s through a near detector (also at Fermilab) onto the \nova{} far detector located in Ash River Minnesota in the USA. For \nova{} we assume $3+\bar{3}$ (three years neutrino and three years antineutrino running) unless specified otherwise. The detector is 14 kt liquid argon detector. Our experimental specification of coincides with \cite{Agarwalla:2012bv}. To perform analysis we use the GLoBES software package along with files for 3+1 case PMNS matrices and probabilities \cite{Huber:2004ka,Huber:2007ji,Huber:2007xx2,Huber:2009xx}.

\section{Identifying new degeneracies in the presence of a sterile neutrino}
\label{sec4}

\begin{table}[tbp]
	\centering
	\setlength{\extrarowheight}{0.1cm}
	\begin{tabular}{|c|c|c|}
		\hline
		$4\nu$ Parameters & True Value & Test Value Range\\
		\hline
		$\sin^2\theta_{12}$ & $0.304$ & $\mathrm{N/A}$\\
		$\sin^22\theta_{13}$ & $0.085$ & $\mathrm{N/A}$\\
		$\theta_{23}^{\mathrm{LO}}$ & $40^\circ$ & $(40^\circ,50^\circ)$\\
		$\theta_{23}^{\mathrm{HO}}$ & $50^\circ$ & $(40^\circ,50^\circ)$\\
		$\sin^2\theta_{14}$ & $0.025$ & $\mathrm{N/A}$\\
		$\sin^2\theta_{24}$ & $0.025$ & $\mathrm{N/A}$\\
		$\theta_{34}$ & $0^\circ$ & $\mathrm{N/A}$\\
		$\delta_{13}$ & $-90^\circ$ & $(-180^\circ,180^\circ)$\\
		$\delta_{14}$ &  $-90^\circ,0^\circ,90^\circ$ & $(-180^\circ,180^\circ)$\\
		$\delta_{34}$ & $0^\circ$ & $\mathrm{N/A}$\\
		$\Delta m^2_{21}$ & $7.5\times10^{-5}\mathrm{eV}^2$ & $\mathrm{N/A}$\\
		$\Delta m^2_{31}$ & $2.475\times10^{-3}\mathrm{eV}^2$ & $(2.2,2.6)\times10^{-3}\mathrm{eV}^2$\\
		$\Delta m^2_{41}$ & $1\mathrm{eV}^2$ & $\mathrm{N/A}$\\
		\hline
	\end{tabular}
	\caption{\label{tab:i} Expanded $4\nu$ parameter true values and test marginalisation ranges, parameters with N/A are not marginalised over.
	\label{SterParam}}
\end{table}

The information for the standard oscillation parameters comes from the global analysis of the world neutrino data \cite{Forero:2014bxa,Esteban:2016qun,Capozzi:2013csa}. For the sterile neutrino parameters $\theta_{14}$, $\theta_{24}$ and $\Delta m^2_{41}$ our best-fit values are consistent with Refs. \cite{Giunti:2011gz,Giunti:2013aea,Kopp:2013vaa,Gariazzo:2017fdh}. We have set $\theta_{34}$ and $\delta_{34}$ to zero throughout our analysis due to them not appearing in the vacuum equation for $P_{\mu e}$. Our choice of the neutrino oscillation parameters are listed in Table \ref{SterParam}.

\subsection{Identifying degeneracies at the probability level}

In this section we will discuss parameter degeneracies in 3+1 case at the probability level. In Fig. \ref{fig:prob} we plot the appearance channel probability $P(\nu_\mu \rightarrow\nu_e)$ vs energy for the NO$\nu$A baseline. For plotting the probabilities we have averaged the rapid oscillations due to $\Delta m^2_{41}$. The left column corresponds to neutrinos and the right column corresponds to antineutrinos. In all the panels $\delta_{13}$ is taken as $-90^\circ$ and the bands are due to the variation of $\delta_{14}$. 

The upper panels of Fig. \ref{fig:prob} shows the hierarchy-$\delta_{14}$ degeneracy. For these panels $\theta_{23}$ is taken as $45^\circ$. NH (IH) corresponds to $\Delta m^2_{31} = + (-)2.4 \times 10^{-3}$ eV$^2$. In both the panels the blue bands correspond to NH and the red bands correspond to IH. Note that in the neutrino probabilities, the green band is above the red band and it is opposite in the antineutrinos. This is because, the matter effect enhances the probability for NH for neutrinos and IH for antineutrinos. For each given band, $\delta_{14} = -90^\circ$ corresponds to the maximum point in the probability and $+90^\circ$ corresponds to the minimum point in the probability, for both neutrinos and antineutrinos. These features in the probability can be understood in the following way. From Eq. \ref{eq1}, we see the neutrino appearance channel probability depends on the phases as: $a + b \cos (\Delta_{31} + \delta_{13}) + c \sin(\Delta_{31} +\delta_{13} - \delta_{14})$, where $a$, $b$ and $c$ are positive quantities. At the oscillation maxima we have $\Delta_{31} = 90^\circ$. As our probability curves correspond to $\delta_{13} = -90^\circ$, for neutrinos we obtain $a + b - c \sin\delta_{14}$. Now it is easy to understand that the contribution to the probability will be maximum for $\delta_{14} = -90^\circ$ and minimum for $\delta_{14} = +90^\circ$. Now let us see what happens for antineutrinos. For antineutrinos, we change sign of $\delta_{13}$ and $\delta_{14}$ in Eq. \ref{eq1} and we obtain for $\delta_{13} = -90^\circ$ as $a - b - c \sin\delta_{14}$. Thus even for the antineutrinos, the probability is maximum for $\delta_{14} = -90^\circ$ and minimum for $\delta_{14} = +90^\circ$. This is in stark contrast to the behaviour of $\delta_{13}$, as in the standard three flavor case, $\delta_{13} = -90^\circ$ corresponds to the maximum probability while $\delta_{13} = +90^\circ$ corresponds the minimum probability for neutrinos (vice-versa for antineutrinos). From the plots we see that there is overlap between \{NH, $\delta_{14} = 90^\circ$\} and \{IH, $\delta_{14} = -90^\circ$\} for the neutrinos and \{NH, $\delta_{14} = -90^\circ$\} and \{IH, $\delta_{14} = +90^\circ$\} for antineutrinos. Thus we understand that unlike the nature of hierarchy-$\delta_{13}$ degeneracy, the hierarchy-$\delta_{14}$ degeneracy is different in neutrinos and antineutrinos so in principle a balanced combination of neutrino and antineutrino should be able to resolve this degeneracy.

In the lower panels of Fig. \ref{fig:prob}, we depict the octant-$\delta_{14}$ degeneracy. In these panels LO corresponds to $\theta_{23} = 40^\circ$ and HO corresponds to $50^\circ$. Here the hierarchy is chosen to be normal with $\Delta m^2_{31} = + 2.4 \times 10^{-3}$ eV$^2$.
In both the panels, the blue band correspond to LO and the red band correspond to HO. Note that in both the panels, the red band is above the blue band. This is because the appearance channel oscillation probability increases with increasing $\theta_{23}$ for both neutrinos and antineutrinos. As already explained in the above paragraph, for each given band, $\delta_{14} = -90^\circ$ corresponds to the maximum value in the probability and $\delta_{14} = +90^\circ$ to the minimum point in the probability for both neutrinos and antineutrinos. From the panels we see that (LO, $\delta_{14} = -90^\circ$) is degenerate with (HO, $\delta_{14} = +90^\circ$). It is interesting to note that this degeneracy is same in both neutrinos and antineutrinos \cite{Agarwalla:2016xlg}. This is a remarkable difference compared to the octant-$\delta_{13}$ degeneracy which is different for neutrinos and antineutrinos. Thus we understand that in the 3+1 scenario, it is impossible to remove the octant degeneracy by combining neutrino and antineutrino runs. 

\begin{figure*}
     \includegraphics[width=0.45\textwidth]{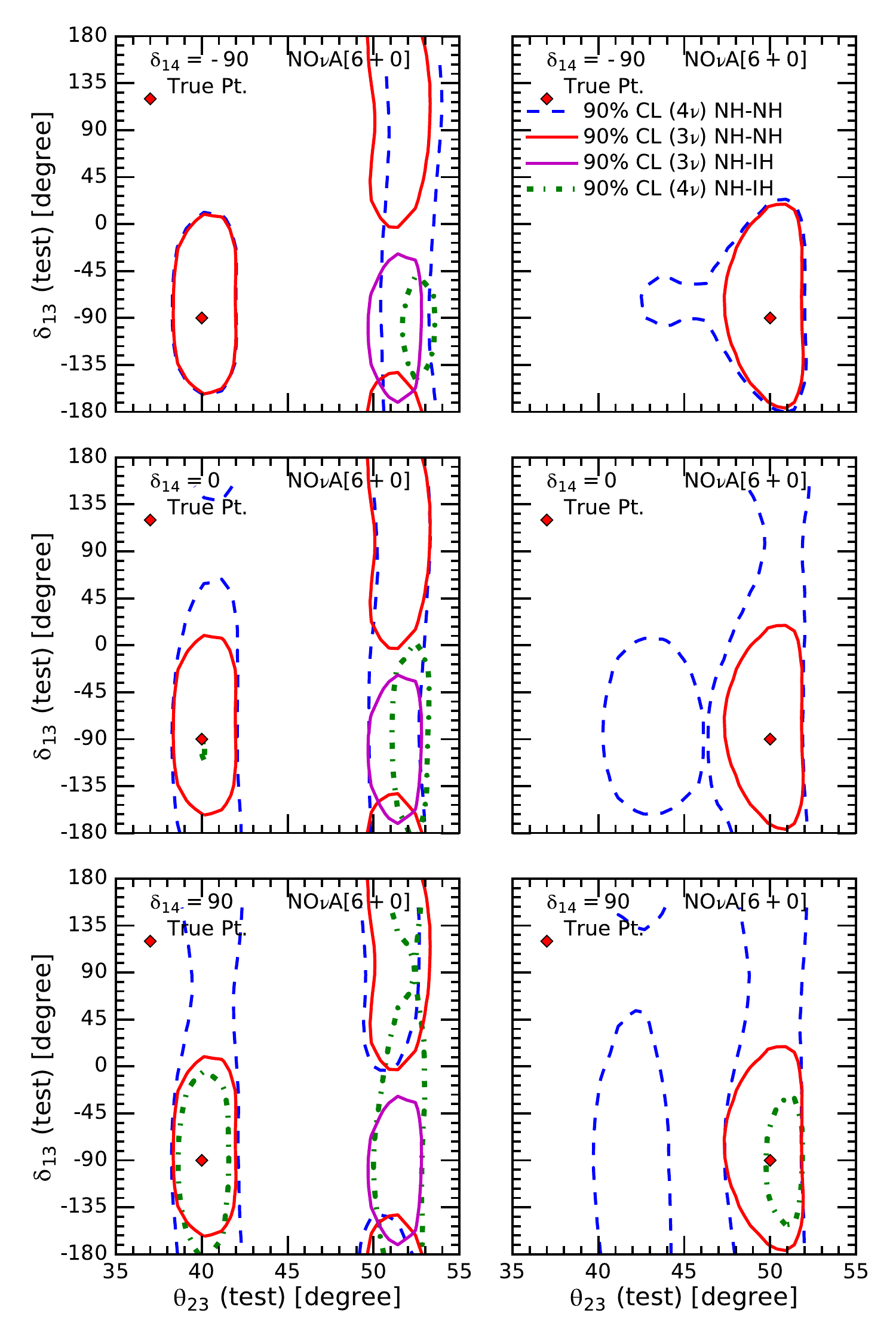}
     \includegraphics[width=0.45\textwidth]{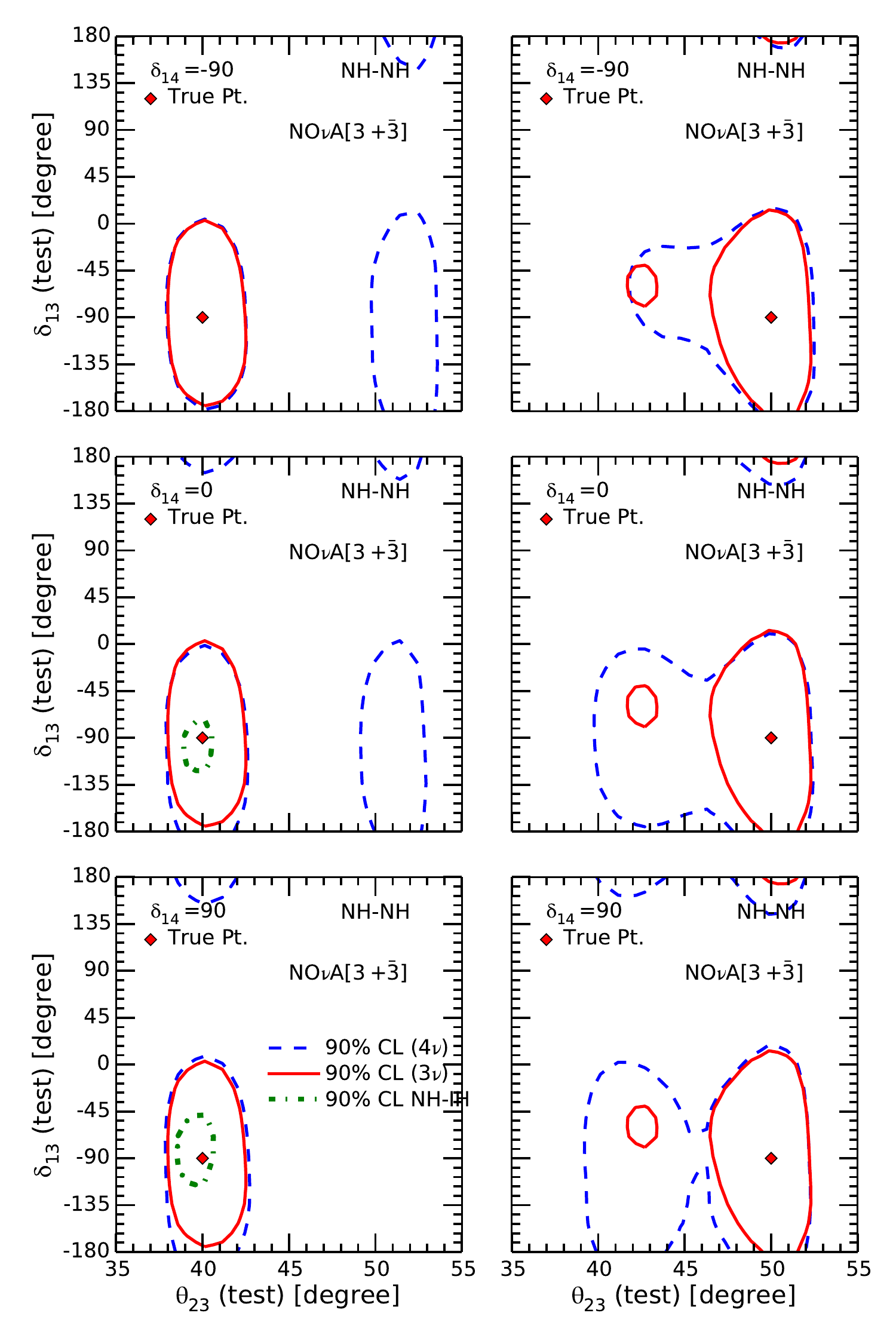}
     \caption{Contour plots in the $\theta_{23}({\rm test})$ vs $\delta_{13}({\rm test})$ plane for two different true values of $\theta_{23}= 40^\circ$ (first and third column) and $50^\circ$ (second and fourth column) for NO$\nu$A $(6+\bar0)$ (first and second column) and ($3+\bar 3$) (third and fourth column). The first, second and third rows are for $\delta_{14}=-90^\circ$ , $0^\circ$ and $90^\circ$ respectively. The true value for the $\delta_{13}$ is taken to be $-90^\circ$. The true hierarchy is NH. We marginalize over the test values of $\delta_{14}$. Also shown is the contours for the $3\nu$ flavor scenario.}
     \label{fig:hierarchy_degen}
\end{figure*}

\subsection{Identifying degeneracies at the event level}

Now we analyze the relevant degeneracies at the $\chi^2$ level. In Fig. \ref{fig:hierarchy_degen} we have given the contours in the $\theta_{23} ({\rm test})$-$\delta_{13} ({\rm test})$ plane for three different values of $\delta_{14}$ at $90\%$ C.L. The first and second column correspond to the case when NO$\nu$A runs in pure neutrino mode and the third and fourth column correspond to the case when NO$\nu$A runs in equal neutrino and antineutrino mode. Note that though the current plan for NO$\nu$A is to run in the equal neutrino and antineutrino mode, we have produced plots corresponding to the pure neutrino run of NO$\nu$A to understand the role of antineutrinos in resolving the degeneracies. We have chosen the true parameter space to coincide with the latest best-fit of NO$\nu$A i.e. $\delta_{13} = -90^\circ$ and NH. While generating the plots we have marginalized over $\delta_{14}$, $|\Delta m^2_{31}|$ in the test parameters while all the other relevant parameters are kept fixed in both the true and test spectrum. The top, middle and bottom rows correspond to $\delta_{14} = -90^\circ$, $0^\circ$ and $+90^\circ$ respectively. In each row the first and third panel correspond to LO ($\theta_{23} = 40^\circ$) and the second and fourth panel correspond to HO ($\theta_{23} = 50^\circ$). These values of $\theta_{23}$ are the closest to the current best-fit according the latest global analysis. For comparison we also have given the contours for the standard three generation case. Note that because of the existence of hierarchy-$\delta_{14}$ and octant-$\delta_{14}$ degeneracies, there will be three spurious solutions in addition to the true solution which are the: (i) right hierarchy-wrong octant (RH-WO), (ii) wrong hierarchy-right octant (WH-RO) and (iii) wrong hierarchy-wrong octant (WH-WO) solutions. As the hierarchy-$\delta_{14}$ and octant-$\delta_{14}$ degeneracy occurs for any given value of $\delta_{13}$ (which is $-90^\circ$ in this case), 
all the above mentioned three spurious solutions should appear at the correct value of $\delta_{13} ({\rm test})=-90^\circ$. Below we discuss the appearance of these spurious solutions in detail.

Let us start with the three generation case. The red contour is for RH solutions and the purple contour is for WH solutions. For NO$\nu$A $(6+\bar0)$ and LO (first column), we see that apart from correct solution (the contour around the true point), there is a RH-WO solution around $\delta_{13} ({\rm test}) = +90^\circ$ and a WH-WO solution around $\delta_{13} ({\rm test}) = -90^\circ$. Note that both of these wrong solutions vanish in the NO$\nu$A $(3+\bar3)$ case (third column). This is because as we mentioned earlier, the octant degeneracy in the standard three flavor scenario behaves differently for neutrinos and antineutrinos and a balanced combination of them can resolve this degeneracy. On the other hand for NO$\nu$A $(6+\bar0)$ and HO (second column), there are no wrong solutions apart from the true solution but in NO$\nu$A $(3+\bar3)$ (fourth column), a small RH-WO solution appears around $\delta_{13} ({\rm test}) = -90^\circ$. This can be understood in the following way. The addition of antineutrinos helps in the sensitivity only if there is degeneracy in the pure neutrino mode. But if there is no degeneracy, then replacing neutrinos with antineutrinos causes a reduction in the statistics as the neutrino cross section is almost three times higher than the
antineutrino cross section. As \{$\delta_{13} = -90^\circ$, NH, HO \} does not suffer from degeneracy in the pure neutrino mode, addition of antineutrinos makes the precision of $\theta_{23}$ worse as compared to NO$\nu$A ($6+\bar0$) and a WO solution appears for NO$\nu$A $(3+\bar3)$.

Now let us discuss the case for the 3+1 scenario for $\delta_{14}=-90^\circ$ (first row). In these figures the blue contours correspond to the RH solution and the green contours correspond to the WH solutions. For NO$\nu$A $(6+\bar0)$ and LO (first panel), we see that there is a RH-WO solution for the entire range of $\delta_{13} ({\rm test})$. Note that NH and $\delta_{14} = -90^\circ$ don't suffer from the hierarchy-$\delta_{14}$ degeneracy but we find a WH solution appears with WO around $\delta_{13} ({\rm test}) = -90^\circ$ which disappears in the NO$\nu$A $(3+\bar3)$ case (third panel). The RH-WO solution around $\delta_{13} ({\rm test}) = -90^\circ$ on the other hand, remains unresolved even in the NO$\nu$A $(3+\bar3)$ case. This is because that the octant - $\delta_{14}$ degeneracy is same for neutrinos and antineutrinos. This is one of the major new features of the 3+1 case when compared to the three generation case. In the three generation case, NO$\nu$A $(3+\bar3)$ is free from all the degeneracies for $\delta_{13} = -90^\circ$ in NH and LO but if we introduce a sterile neutrino, then there will be an additional WO solution even at 90\% C.L. For HO, we see that $(6+\bar0)$ configuration is almost free from any degeneracies except for a small RH-WO solution (second panel). For NO$\nu$A $(3+\bar3)$, the lack of statistics decrease the $\theta_{23}$ precision and there is a growth in the WO region (fourth panel).

Next let us discuss the case for $\delta_{14}=+90^\circ$ (third row). For $6+\bar0$ and LO (first panel) we see that there is a WH-RO solution around $\delta_{13}({\rm test}) = -90^\circ$, a WH-WO solution for the entire range of $\delta_{13}({\rm test})$ and RH-WO solution around 
$\delta_{13}({\rm test}) = +90^\circ$. In this case the inclusion of the antineutrino run of NO$\nu$A (third panel) almost resolves all the degenerate solutions but a small WH solution remains unresolved. This indicates that in this case the statistics of the antineutrino run are not sufficient to remove the RH-WO solution. For HO, we have the RH-WO and WH-RO solutions, both at $\delta_{13} ({\rm test}) = -90^\circ$ for NO$\nu$A($6+\bar0$)  (second panel). For NO$\nu$A($3+\bar3$) we see that the WH solution gets removed but the WO solution remains unresolved (fourth panel).

For $\delta_{14} = 0^\circ$ (middle row), we see that there is a RH-WO solution in the entire range of $\delta_{13} ({\rm test})$  and WH-WO solution around $\delta_{13} ({\rm test}) = -90^\circ$ for NO$\nu$A $(6+\bar0)$ in LO (first panel). By the inclusion of antineutrino run, the WH-WO region gets resolved but the RH-WO solution remains unresolved at $\delta_{13} ({\rm test}) = -90^\circ$  (third panel). Apart from that, there is also  the emergence of a WH-RO solution at $\delta_{13} ({\rm test}) = -90^\circ$. For the HO, we see that apart from the true solution, there is a RH-WO region for both NO$\nu$A $(6+\bar0)$ and $(3+\bar3)$ configurations around at $\delta_{13} ({\rm test}) = -90^\circ$ (second and fourth panel respectively).

\begin{figure}
     \includegraphics[width=0.45\textwidth]{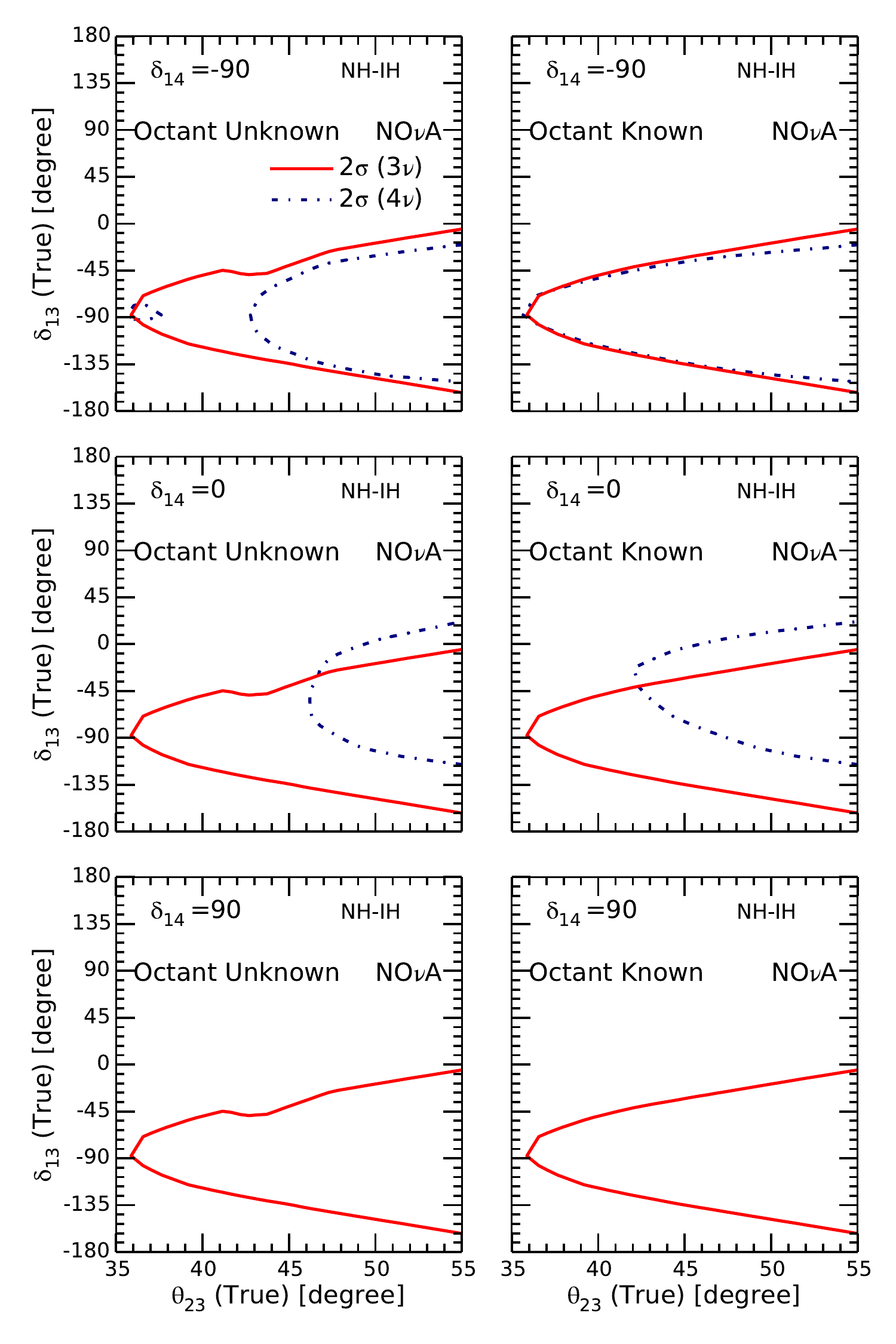}
     \caption{Contour plots at $2\sigma$ C.L. in the $\theta_{23}({\rm true})$ vs $\delta_{13}({\rm true})$ plane for Octant Unknown (left panel) and Octant Known (right panel) scenarios for NO$\nu$A ($3+\bar 3$). The first, second and third rows are for $\delta_{14}=-90^\circ$, $0^\circ$ and $90^\circ$ respectively. The true and test hierarchies are chosen to be normal (NH) and inverted hierarchy (IH) respectively. Also shown contours for the $3\nu$ flavor scenario.}
     \label{fig:hierarchy_degen_1}
\end{figure}

\section{Results for hierarchy sensitivity}
\label{sec5}

We now discuss the hierarchy sensitivity of NO$\nu$A $(3+\bar3)$ in the presence of a sterile neutrino. In the Fig. \ref{fig:hierarchy_degen_1} we have given the $2\sigma$ hierarchy contours in the $\delta_{14}({\rm true})$ - $\theta_{23}({\rm true})$ plane for three values of $\delta_{14}$. 
The red contours are for standard three flavor case and the blue contours are for 3+1 case. For the region inside the contours one can exclude the wrong hierarchy at $2\sigma$. Here the true hierarchy is NH. While generating these plots we have marginalized over test values of $\delta_{13}$, $\delta_{14}$ and $|\Delta m^2_{31}|$. We have assumed the octant to be unknown and known in the left and right panels respectively. The top, middle and bottom rows corresponds to $\delta_{14} = -90^\circ$, $0^\circ$ and $90^\circ$ respectively. 

For the standard three flavor scenario we see NO$\nu$A has $2\sigma$ hierarchy sensitivity around $-90^\circ$ for all the values of $\theta_{23}$ ranging from $35^\circ$ to $55^\circ$. This is irrespective of the information of the octant. This is because for NO$\nu$A $(3+\bar3)$, $\delta_{13} = -90^\circ$ do not suffer from hierarchy degeneracy in NH. This can be understood from Fig. \ref{fig:hierarchy_degen} by noting the absence of purple contour in NO$\nu$A $(3+\bar3)$ for both LO and HO. 

In the 3+1 case, if $\delta_{14}$ is $-90^\circ$ then the hierarchy sensitivity is lost when $\theta_{23}$ is less than $43^\circ$ in the known octant case (top left panel). Note that though NO$\nu$A $(3+\bar3)$ does not have a WH solution at 90\%, the loss of hierarchy sensitivity implies that this degeneracy re-appears at $2\sigma$. If the octant is known then the sensitivity of 3+1 coincides with the standard 3 flavor case (top right panel). This signifies that the loss of sensitivity in the 3+1 case for the value of $\delta_{14} = -90^\circ$ is mainly due to the WH-WO solution. In the middle row we see that in the 3+1 case, one cannot have hierarchy sensitivity at $2\sigma$ for true $\delta_{14} = 0^\circ$ if $\theta_{23}$ is less than $46^\circ$ ($42^\circ$) when the octant is unknown (known) as can be seen from the middle panels. This implies that for this value of true $\delta_{14}$ the hierarchy sensitivity of NO$\nu$A is affected by the WH solution occurring with both right and wrong octant. 
But the most remarkable result is found for $\delta_{14} = 90^\circ$ (bottom panels). For this value of $\delta_{14}$ we see that the hierarchy sensitivity of NO$\nu$A is completely lost. This is mainly due to the WH-RO solution. Thus we understand that if there exists a $\sim1\mathrm{eV}$ sterile neutrino in addition to the three active neutrinos and the value of $\delta_{14}$ chosen by nature is $+90^\circ$, then NO$\nu$A can not have even a $2\sigma$ hierarchy sensitivity for $\delta_{13} = -90^\circ$ and NH which is present best fit of NO$\nu$A.

\section{Conclusion}
\label{sec6}

In this work we have studied the parameter degeneracy and hierarchy sensitivity of NO$\nu$A in the presence of a sterile neutrino. Apart from the hierarchy-$\delta_{13}$ and octant-$\delta_{13}$ degeneracy in the standard three flavor scenario, we have identified two new degeneracies appearing with the new phase $\delta_{14}$ which occur for every value of $\delta_{13}$. These are hierarchy-$\delta_{14}$ degeneracy and octant-$\delta_{14}$ degeneracy. Unlike the standard three generation case, here the octant degeneracy behaves similarly for neutrinos and
antineutrinos and the hierarchy degeneracy behaves differently. Thus a combination of neutrinos and antineutrinos are unable to resolve the octant-$\delta_{14}$ degeneracy but can resolve the hierarchy-$\delta_{14}$ degeneracy. To identify the degenerate parameter space we present our results in $\theta_{23}({\rm test})$ - $\delta_{13}({\rm test})$ plane for three values of $\delta_{14}({\rm true})$ assuming (i) NO$\nu$A runs in pure neutrino mode and (ii) NO$\nu$A runs in equal neutrino and antineutrino mode. We have chosen normal hierarchy and $\delta_{13} = -90^\circ$ motivated by the latest fit from NO$\nu$A data. In those plots we find that there are different RH-WO, WH-RO and WH-WO regions depending on the 
true nature of the octant of $\theta_{23}$ and true value of $\delta_{14}$. From these plots we find that the addition of antineutrinos helps to resolve the WH solutions but fails to remove the WO solutions appearing at $\delta_{13} ({\rm test}) = -90^\circ$. However we find that for $\delta_{14} ({\rm true}) = 90^\circ$ and LO, the antineutrino run of NO$\nu$A is unable to resolve the WH solution appearing with right octant at 90\% C.L. While for $\delta_{14} ({\rm true}) = 0^\circ$, the WH-RO solution grows in size for NO$\nu$A $(3+\bar3)$ as compared to NO$\nu$A $(6+\bar0)$. Comparing these with that of standard three flavor case we find that apart from the small RH-WO regions for the true higher octant, there are no other degenerate allowed regions for this choice of $\delta_{13}({\rm true})$ and hierarchy in the three flavor case for NO$\nu$A $(3+\bar3)$. Note the region $\delta_{13} = -90^\circ$ and NH is the favorable parameter space of NO$\nu$A which does not suffer from hierarchy-$\delta_{13}$ degeneracy in the standard three flavor scenario where NO$\nu$A can have good hierarchy  sensitivity. But now in the 3+1 case, the hierarchy sensitivity of NO$\nu$A for this parameter value can suffer due to the existence of the new degeneracies. To study that we plot the $2\sigma$ hierarchy contours in the $\theta_{23}({\rm true})$-$\delta_{13}({\rm true})$ plane for three values of true $\delta_{14}$ in NH. While in the standard three flavor case one can have $2\sigma$ hierarchy sensitivity for all the values of $\theta_{23}$ ranging from $35^\circ$ to $55^\circ$, in the 3+1 case we find that for $\delta_{14} = -90^\circ$ and $\theta_{23} = 43^\circ$ the hierarchy sensitivity of NO$\nu$A is lost. For the value of $\delta_{14} = 0^\circ$, the hierarchy sensitivity of NO$\nu$A is also compromised if $\theta_{23}$ is less than $46^\circ$. But the most serious deterioration in hierarchy sensitivity occurs if the value of $\delta_{14}$ chosen by nature is $+90^\circ$. 
At this value of $\delta_{14}$, NO$\nu$A suffers from hierarchy degeneracy and thus it has no hierarchy sensitivity for any value of $\theta_{23}$. Therefore if: (i) the hint of $\delta_{13} = -90^\circ$ persists; (ii) the data begins to show a preference towards LO; and (iii) the observed hierarchy sensitivity is less than the expected sensitivity, then this can be a signal from NO$\nu$A towards existence of a sterile neutrino with $\delta_{14} \neq -90^\circ$. 

\section{Acknowledgement}

The work of MG is supported by the ``Grant-in-Aid for Scientific Research of the Ministry of Education, Science and Culture, Japan", under Grant No. 25105009. SG, ZM, PS and AGW acknowledge the support by the University of Adelaide and the Australian Research Council through the ARC Centre of Excellence for Particle Physics at the Terascale (CoEPP) (grant no. CE110001004).

\end{document}